%% file: OPTICAmeetings.tex
\documentclass[letterpaper,10pt]{article} 

\usepackage{opticameet3} 

\usepackage{amsmath,amssymb}
\usepackage{mathrsfs}
\usepackage{pgfplots} 
\pgfplotsset{compat=1.16}
\usepgfplotslibrary{colorbrewer}
\usetikzlibrary{external}
\usepackage[colorlinks=true,bookmarks=false,citecolor=blue,urlcolor=blue]{hyperref} 

\begin{document}

\title{Optical arbitrary waveform generation using spectro-temporal unitary transforms}


\author{Callum Deakin}

\address{Nokia Bell Labs, 600 Mountain Ave, Murray Hill, NJ, USA}

\email{callum.deakin@nokia-bell-labs.com} 


\begin{abstract}
We discuss the prospect of using cascaded phase modulators and dispersive elements to achieve arbitrary optical waveform generation. This transform is not limited by the bandwidth of its constituent modulators and is theoretically lossless.
\end{abstract}
\vspace{0.3cm}

\section{Introduction}

Coherent modulation allows for full, arbitrary manipulation of the optical carrier and enables efficient digital signal processing for compensation of link and transceiver distortions in optical networks. Usually this is achieved by using two 90 degree out of phase Mach-Zehnder amplitude modulators (MZMs) for each orthogonal polarization, which allows for a linear transfer of a complex radio-frequency (RF) signal (e.g. QAM) to the optical carrier. This arrangement, show in in Fig.~\ref{experiment_setup}(a), is constrained in bandwidth by the bandwidth of its constituent phase modulators and their corresponding analog drivers and DACs. Furthermore, it is fundamentally lossy since the optical power in low amplitude symbols is dumped into the terminated output port of the MZMs, resulting in modulation losses that often exceed 25~dB in modern coherent transmitters.

In this paper we discuss an alternative optical arbitrary waveform technique that can be used to generate coherent optical signals. The technique uses a cascade of alternating phase modulators and dispersive elements (shown in Fig.~\ref{experiment_setup}(b)) to achieve arbitrary waveform generation. Importantly, is not constrained by the bandwidth of its constituent modulators and is theoretically lossless, since it relies only on temporal and spectral phase modulations to temporally redistribute light from the input laser and achieve the desired waveform. 

We review the theoretical principles that allow these transforms as well as the optimisation procedures required to find the correct phase instructions. We also discuss the schemes complexity compared to coherent modulation and practical considerations when implementing this in real time. Finally, we show some example transformations and discuss the required components to implement these on-chip.

\section{Spectro-temporal unitary transformations}
Any linear optical transformation can described by a unitary matrix operator $U \in  \mathscr{U}_k(\mathbb{C})$, which transforms an vector describing $k$ input modes into a vector of $k$ output modes. To achieve arbitrary transformation between the input and output modes, we must be able to construct every unitary matrix. It has been proven that any unitary matrix can be written as a product of unitary diagonal matrices and a matrix $H$ which has all its entries non-zero~\cite[Proposition 7]{schmid2000decomposing}, which is an extension of~\cite[Lemmas 2 and 3]{borevich1981subgroups}. Therefore, we can construct any unitary matrix with a succession of diagonal matrices $\Lambda_n$ and a fixed matrix $H$ which has no zero entries 
\begin{equation}
    U = \Lambda_1H\Lambda_2H\Lambda_3H \cdots \Lambda_nH
\end{equation}

The physical interpretation of $H$ is a mode-mixing operation: every output mode is a linear composition involving every input mode. An example of a matrix which fulfils this condition is the discrete Fourier transform matrix
\begin{equation}
    \textnormal{DFT}_m = \frac{1}{\sqrt{m}}\Big(\exp{\frac{2\pi i}{m} \Big)}^{j\cdot k}_{j,k=0,\dots, m-1}
\end{equation}

for an $m\times m$ matrix. We can therefore build any mode transformation using a product of diagonal matrices and the discrete Fourier transform matrix~\cite{morizur2010programmable}. 
\begin{figure*}[tb]
   \centering
    \includegraphics[width=0.8\linewidth]{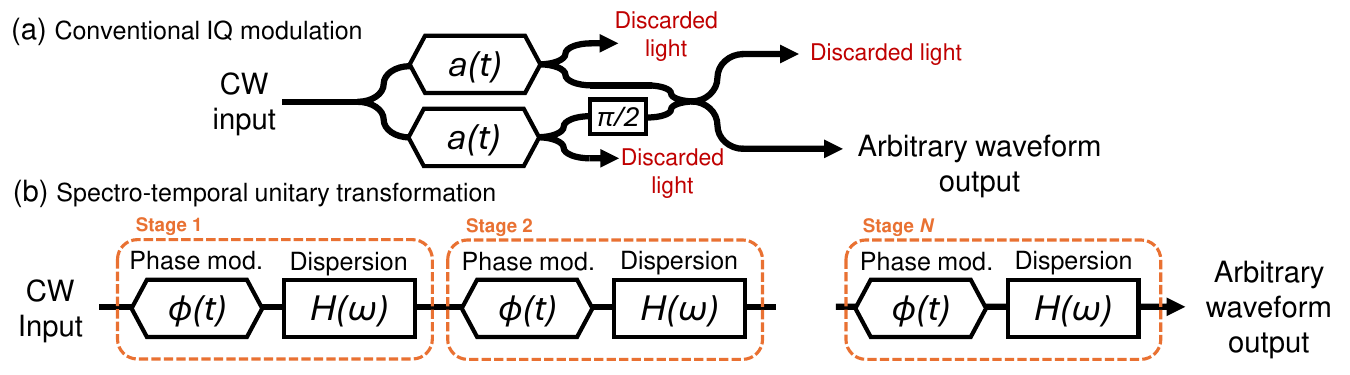}
    \caption{(a) Conventional IQ modulation based on amplitude modulation, $a(t)$. (b) Lossless spectro-temporal unitary transform based arbitrary waveform modulation based on cascaded phase modulators $\phi(t)$ and dispersive elements $H(\omega)$.}
    \vspace{-0.5cm}
    \label{experiment_setup}
\end{figure*}
Implementing a linear optical transformation based on this composition requires two optical components: a phase modulator that modulates the optical modes independently (diagonal matrix), and a device that replicates the mode mixing operation of the DFT. These two devices represent a single stage of the transform, which can then be cascaded together in $N$ stages, as shown in Fig.~\ref{experiment_setup}(b).

This method was first described and implemented on spatial modes~\cite{morizur2010programmable}, in which the modulator is a spatial light modulator. The DFT matrix on the other hand is approximated by free space propagation, due to the well known fact that the Fraunhofer (far-field) diffraction pattern approximates a Fourier transform on the spatial modes. This spatial mode converter is known as a multi-plane light wave converter (MPLC). As an example, MPLCs can be used as mode multiplexers to convert spatially separated Gaussian spots into spatially overlapped orthogonal mode sets, e.g. for propagation in a multimode optical fiber or a free space optical link~\cite{fontaine2019laguerre}.

Noting the well known duality between the paraxial diffraction of spatial modes and the narrowband dispersion of temporal modes~\cite{thiel2017programmable,kolner1994space}, this spatial mode converter concept was replicated in the temporal domain by replacing the spatial light modulator with a temporal phase modulator, and free space propagation with a dispersive element~\cite{mazur2019optical}. This arrangement is known as a spectro-temporal unitary transform because it relies on successive transforms and modulations between the light fields spectral and temporal modes, enabled by the Fourier transform property of the dispersive element. This principle has previously been applied to ultrafast pulse shaping~\cite{azana2005reconfigurable,salem2013application}, and it was also shown that it can modulate individual lines of a frequency comb independently, removing the requirement for mux/demux devices in multi-channel systems~\cite{mazur2019multi}. 

\section{Calculation of phase instructions and complexity}

Although we have shown that in general any waveform can be generated by a cascade of phase modulators and dispersion estimates, finding the set of phase instructions $\{\phi_1(t),\phi_2(t),\dots \phi_N(t)\}$ that will generate a specific waveform is non-trivial and analytic constructions are known for only low mode/phase mask counts~\cite{yasir2025compactifying}. Instead, the phase instructions must be determined by a optimization procedure that minimises the noise-to-signal ratio (NSR) compared to the target waveform $\Psi_\textnormal{target}(t)$. The NSR can be calculated as~\cite{saxena2023performance}
\begin{equation} \label{NSR_unconstrained}
\textnormal{NSR} = \frac{1}{E_\textnormal{target}}\sum_{m=0}^M |\Psi_\textnormal{target}(mT) - F_N(mT)|^2
\end{equation}
where $F_N(t)$ is the forward propagating wave after $N$ stages, $E_\textnormal{target}$ is the energy of the target waveform and $T$ is the sampling period. Since the gradient with respect to the phase instructions can also be calculated as
\begin{equation} \label{gradient_unconstrained}
    \frac{\partial \textnormal{NSR}}{\partial \phi_n} =  \frac{2}{E_\textnormal{target}}\sum_{m=0}^M \textnormal{Re} (iF_n^*(mT)B_n(mT))
\end{equation}
for backward propagating wave $B_n(t)$, the phase instructions $\{\phi_1(t),\phi_2(t),\dots \phi_N(t)\}$ can be optimised by using any number of gradient-descent based algorithms, such as the Broyden–Fletcher–Goldfarb–Shanno algorithm (BFGS). In the first demonstration of the spectro-temporal unitary transform technique~\cite{mazur2019optical}, a wavefront-matching algorithm~\cite{sakamaki2007new} was used as in the spatial MPLC formulation~\cite{fontaine2019laguerre}. However, this technique is much less efficient than gradient-descent algorithms~\cite{saxena2023performance} and prevents the use of additional constraints.

In order to implement this technique in a coherent transmitter, the aforementioned optimisation procedure must be calculated in real time and on a continuous stream of target symbols. For continuous operation, an incoming block of symbols is padded with a portion of the neighbouring symbols from the preceding and following block~\cite{saxena2023performance}. The optimisation procedure is then performed on the padded target waveform, with the resulting phase instructions cut to remove the padding before being send to the phase modulators. This ensures that no NSR degradation is observed at the block boundaries and the modulation can be applied in real time.

We must also consider the complexity of the optimisation procedure and whether it is feasible to perform this continuously on an application specific integrated circuit (ASIC). Firstly, we must calculate the NSR function (\ref{NSR_unconstrained}) by calculating the forward propagating wave and comparing it to the target waveform. This requires four operations at every stage: 1) phase modulation (complex multiplication), 2) FFT, 3) dispersion (complex multiplication) and 4) inverse FFT. Given that a single FFT requires $\frac{K}{4} \log_2(K)$ complex multiplications for a block length $K$~\cite{spinnler2009complexity}, calculation of (\ref{NSR_unconstrained}) requires $N (\frac{K}{2} \log_2(K) + 2K)$ complex multiplications for $N$ stages. In addition, the gradient function requires calculation of the backwards propagating wave which has the same complexity. Both of these quantities need to be calculated at every iteration of the optimization procedure, leading to a total of $NK(\log_2(K) + 4)$ complex multiplications per iteration. This is $2N$ the complexity of a simple frequency domain pulse shaping filter, and is only the minimum complexity of a single iteration. 

The actual complexity in terms of operations per symbols depends on the target NSR, number of stages, and optimisation algorithm but the most efficient algorithms result in around 2500 times more operations per symbol than conventional coherent IQ modulation~\cite{saxena2023performance}. This highlights that computation complexity is a significant drawback of this scheme currently. Nonetheless, significant progress has been made in recent years to improve the optimisation efficiency, including deterministic algorithms~\cite{huhtanen2015factoring,lopez2021arbitrary} and instead minimising the geodesic distance on the unitary group manifold as opposed to the simple Euclidean distance to the target waveform~\cite{alvarez2025universality}.

Notwithstanding any research efforts to reduce this complexity, the increasing parallelisation and efficiency of advanced CMOS nodes may make the scheme attractive by trading computational complexity with the improved optical performance. On the other hand, the technique may benefit applications that do not require continuous operation and are therefore not impacted by this complexity, such as test and measurement or temporal mode transformations for ultrafast or quantum applications~\cite{ashby2020temporal}.

\section{Example transformation: RRC shaped 64-QAM}

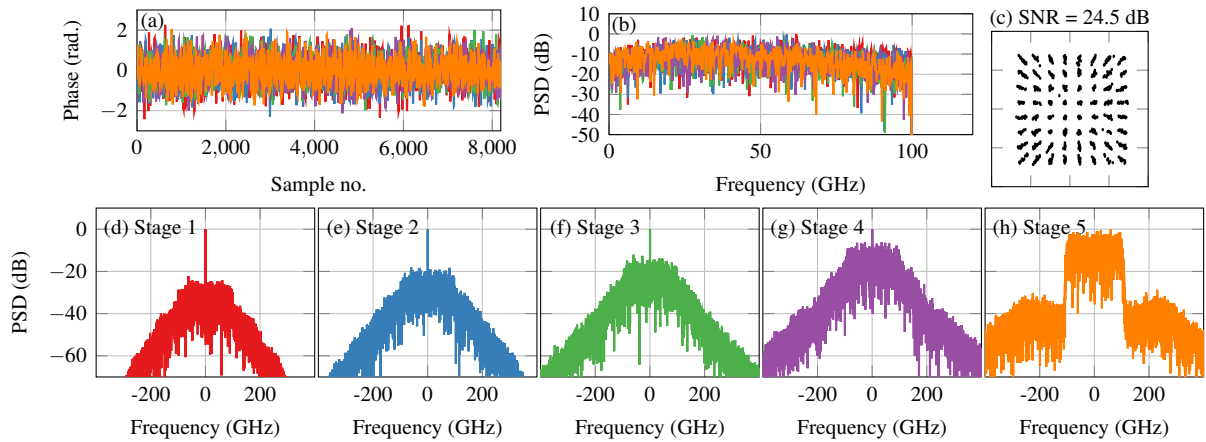
\begin{figure}[t]
   \centering
    \input{figs/fig_spectrum_compiled}
    \caption{Example 5 stage transformation for generating a block of 1024 RRC-shaped 200~GBd 64-QAM symbols using 100~GHz phase modulator bandwidth and 20~ps/nm dispersion per stage, sampled at 1600~GSa/s.  (a) Phase instructions time series. (b) Phase instruction spectra. (c) Generated constellation. (d)-(h) Optical spectrum after each stage.}
    \label{fig:example}
    \vspace{-0.5cm}
\end{figure}

An example simulated transformation is shown in Fig.~\ref{fig:example}, generating a block of 1024 RRC-shaped 200~GBd 64-QAM symbols using 100~GHz phase modulator bandwidth and 20~ps/nm dispersion per stage, sampled at 1600~GSa/s. The phase instructions are calculated using the limited memory BFGS algorithm. This amount of dispersion is available on chip using ring or grating structures~\cite{stern2023silicon}. Importantly, the amount of dispersion required decreases quadratically with increasing baudrate, relaxing this requirement for future higher baudrate systems~\cite{deakinofc2025}.

\bibliographystyle{opticajnl}
\bibliography{sample}

\end{document}

%% file: figs/fig_spectrum_compiled.tex
\includegraphics[]{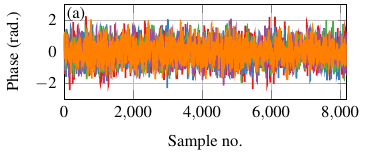}\includegraphics[]{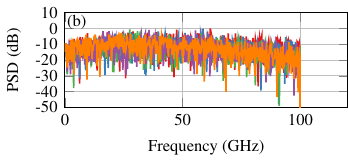}\includegraphics[]{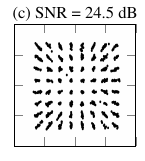}

\includegraphics[]{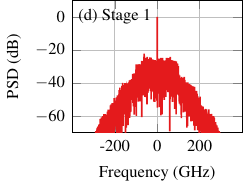}\hspace{=-0.2cm}\includegraphics[]{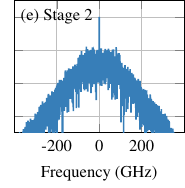}\hspace{=-0.2cm}\includegraphics[]{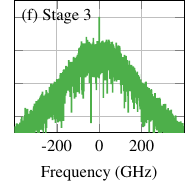}\hspace{=-0.2cm}\includegraphics[]{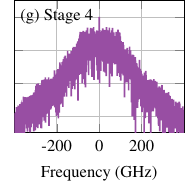}\hspace{=-0.2cm}\includegraphics[]{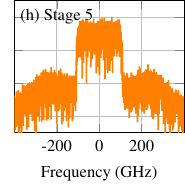}

%% file: sample.bib
@article{kolner1994space,
  title={Space-time duality and the theory of temporal imaging},
  author={Kolner, Brian H},
  journal={IEEE Journal of Quantum Electronics},
  volume={30},
  number={8},
  pages={1951--1963},
  year={1994},
  publisher={IEEE}
}

@article{mazur2019multi,
  title={Multi-wavelength arbitrary waveform generation through...},
  author={Mazur, Mikael and others},
  journal={arXiv preprint arXiv:1907.02595},
  year={2019},
    
}

@inproceedings{stern2023silicon,
  title={Silicon photonic direct-detection phase retrieval receiver},
  author={Stern, Brian and others},
  booktitle={ECOC 2023},
  pages={M.A.2.2},
  year={2023},
}

@article{lopez2021arbitrary,
  title={Arbitrary optical wave evolution with Fourier transforms and phase...},
  author={L{\'o}pez Pastor, V{\'\i}ctor and others},
  journal={Optics Express},
  volume={29},
  number={23},
  pages={38441--38450},
  year={2021},
  publisher={Optical Society of America}
}

@article{huhtanen2015factoring,
  title={Factoring matrices into the product of circulant and diagonal...},
  author={Huhtanen, Marko and others},
  journal={Journal of Fourier Analysis and Applications},
  volume={21},
  number={5},
  pages={1018--1033},
  year={2015},
  publisher={Springer}
}

@article{alvarez2025universality,
  title={Universality and Optimal Architectures for Layered Programmable...},
  author={{\'A}lvarez-Vizoso, Javier and others},
  journal={arXiv preprint arXiv:2510.19397},
  year={2025}
}

@article{yasir2025compactifying,
  title={Compactifying linear optical unitaries using multiport beamsplitters},
  author={Yasir, PA and others},
  journal={arXiv preprint arXiv:2505.11371},
  year={2025}
}

@article{salem2013application,
  title={Application of space--time duality to ultrahigh-speed optical signal...},
  author={Salem, Reza and others},
  journal={Advances in Optics and Photonics},
  volume={5},
  number={3},
  pages={274--317},
  year={2013},
  publisher={Optical Society of America}
}

@inproceedings{spinnler2009complexity,
  title={Complexity of algorithms for digital coherent receivers},
  author={Spinnler, B},
  booktitle={ECOC},
  pages={7.3.6},
  year={2009},
  organization={IEEE}
}

@inproceedings{deakinofc2025,

  author={Deakin, Callum. and others},

  booktitle={OFC}, 
  title={Performance Limits of Spectro-Temporal Unitary Transformations for Coherent...}, 
  year={2025},
  pages={ Tu3C.1 },
  doi={10.1364/OFC.2025.Tu3C.1}}

@article{schmid2000decomposing,
  title={Decomposing a matrix into circulant and diagonal factors},
  author={Schmid, Michael and others},
  journal={Linear Algebra and its Applications},
  volume={306},
  number={1-3},
  pages={131--143},
  year={2000},
  publisher={Elsevier}
}

@article{azana2005reconfigurable,
  title={Reconfigurable generation of high-repetition-rate optical pulse sequences based...},
  author={Aza{\~n}a, Jos{\'e} and others},
  journal={Optics letters},
  volume={30},
  number={23},
  pages={3228--3230},
  year={2005},
  publisher={Optical Society of America}
}

@article{morizur2010programmable,
  title={Programmable unitary spatial mode manipulation},
  author={Morizur, Jean-Fran{\c{c}}ois and others},
  journal={JOSA A},
  volume={27},
  number={11},
  pages={2524--2531},
  year={2010},
  publisher={Optica Publishing Group}
}

@article{sakamaki2007new,
  title={New optical waveguide design based on wavefront matching method},
  author={Sakamaki, Yohei and others},
  journal={JLT},
  volume={25},
  number={11},
  pages={3511--3518},
  year={2007},
  publisher={IEEE}
}

@article{saxena2023performance,
  title={Performance Comparison Between All-Pass and {IQ} Optical Modulation},
  author={Saxena, B and others},
  journal={Journal of Lightwave Technology},
  year={2023},
  publisher={IEEE},
  volume={42},
  number={1},
  pages={201-207},
}

@article{borevich1981subgroups,
  title={Subgroups of the unitary group that contain the group of diagonal matrices},
  author={Borevich, ZI and others},
  journal={Journal of Soviet Mathematics},
  volume={17},
  pages={1951--1959},
  year={1981},
  publisher={Springer}
}

@inproceedings{mazur2019optical,
  title={Optical arbitrary waveform generator based on time-domain multiplane light conversion},
  author={Mazur, Mikael and others},
  booktitle={OFC},
  pages={M1B.3},
  year={2019},
}

@article{ashby2020temporal,
  title={Temporal mode transformations by sequential time and frequency phase...},
  author={Ashby, James and others},
  journal={Optics Express},
  volume={28},
  number={25},
  pages={38376--38389},
  year={2020},
  publisher={Optica Publishing Group}
}

@inproceedings{thiel2017programmable,
  title={Programmable unitary transformation of spectro-temporal modes},
  author={Thiel, Val{\'e}rian and others},
  booktitle={Laser Science},
  pages={JW4A--5},
  year={2017},
}

@article{fontaine2019laguerre,
  title={Laguerre-Gaussian mode sorter},
  author={Fontaine, Nicolas K and others},
  journal={Nature communications},
  volume={10},
  number={1},
  pages={1865},
  year={2019},
  publisher={Nature Publishing Group UK London}
}
